\documentclass[twocolumn,aps]{revtex4}
\usepackage{amsmath}
\usepackage{graphicx}

\newcommand{\nl}{\nonumber \\}

\newcommand{\be}{\begin{equation}}
\newcommand{\ee}{\end{equation}}
\newcommand{\bea}{\begin{eqnarray}}
\newcommand{\eea}{\end{eqnarray}}

\newcommand{\Eq}[1]{Eq.\,(\ref{#1})}

\begin{document}

\title{ Kinetic study for hopping conduction through DNA molecules }
\author{Yong-Gang Yang$^{1}$, Peng-Gang Yin$^{1}$,
       Xin-Qi Li$^{1,2}$\footnote{Corresponding author. E-mail: xqli@red.semi.ac.cn}
       and YiJing Yan$^{2}$}
\address{$^{1}$Institute of Semiconductors,
         Chinese Academy of Sciences, P.O.~Box 912, Beijing 100083, China}
\address{$^{2}$Department of Chemistry, Hong Kong University of Science
         and Technology, Kowloon, Hong Kong }

\date{\today}

\begin{abstract}
Recent experiments indicated that disorder effect in DNA
may lead to a transition of the charge transport mechanism from
band resonant tunnelling to thermal activated hopping.
In this letter, based on Mott's variable-range hopping theory we present
a kinetic study for the charge transport properties of DNA molecules.
Beyond the conventional argument in large-scale systems,
our numerical study for finite-size
DNA molecules reveals a number of unique features for
(i) the I-V characteristics, (ii) the temperature and length dependence,
and (iii) the transition from conducting to insulating behaviors.\\
\\
PACS numers: 87.14.Gg,72.20.Ee,72.80.Le
\end{abstract}

\maketitle




The fundamental electronic process in DNA molecules
(DNA electronics) has received great interest in recent years.
In addition to a large number of {\it indirect}
optical measurements, recent {\it direct} electrical measurements for
charge transport through DNA molecules
revealed amazingly conflicting transport behaviors, ranging from
insulator\cite{Bra98,Pab00,Dek01,Ong02},
Ohmic conductor \cite{Fin99,Cai00,Hwa02},
semiconductor \cite{Por00}, to even superconductor \cite{Kas01}.
As a consequence, the intrinsic charge migration mechanism
remains highly controversial.

While quantum transport (in terms of band resonant tunnelling)
picture \cite{Hjo01,Li01,Dek02} was attributed to the
observed conductor/semiconductor behaviors,
Mott's variable range hopping (VRH) theory was proposed
to understand the temperature dependence of the optically
measured conductivity of $\lambda$-DNA \cite{Yu01,Tra00}.
It was also suggested that it is the disorder effect leading to
the insulating behavior observed in other experiments
\cite{Pab00,Dek01,Ong02}.
In particular, the experiment by Yoo {\it et al} \cite{Yoo01} provided
clear evidence for a polaron hopping mechanism which is responsible to
the electrical conduction through a DNA molecule with length
about 20nm and containing identical base pairs.
In this Letter, we employ an extended version of the VRH model
to study the hopping conduction through DNA molecules.
Differing from the conventional
treatment in bulk or large scale systems based on a qualitative
argument, we base our analysis on direct numerical simulation
for finite systems,
which is of particular interest in light of recent experiments
\cite{Dek01,Yoo01}.

DNA molecule with random base pair sequence such as the
$\lambda$-DNA \cite{Tra00,Pab00,Ong02}, or, with identical base pairs
but influenced by complex environment \cite{Dek01,Yoo01},
can be treated as a one-dimensional disordered system,
where the dominant channel for charge migration
is a series of hops between the localized states.
The thermal activated hopping rate between two localized states, say,
the $m$th and $n$th states separated by a distance $R_{mn}$,
can be described as \cite{Mot71},
$k_{mn}=k_{0}e^{-2\alpha R_{mn}-W_{mn}/k_{B} T}$,
where $k_{0}$ is the attempt-to-escape rate,
$\alpha ^{-1}$ the localization length, and $W_{mn}$ the energy
difference of the two states.
We identify $W_{mn}=\Delta a/R_{mn}$, where $\Delta $ denotes the
total energy disorder strength of the system,
and $a$ is the distance
between two adjacent base pairs.
Two additional remarks in relation to the hopping model to be adopted
are as follows:
(i) Instead of considering only the most probable hops as in
the standard VRH theory,
the present work will take into account all possible hops with
the probabilities described by $k_{mn}$.
(ii)
The localization length $\alpha $ is to be influenced by the structural
fluctuations (i.e. the dynamic disorder effect); thus it depends on
temperature. Following Yu and Song \cite{Yu01}, we model this effect by
$\alpha =\alpha _{0}+\alpha _{1}\tanh (T/T_{d})^{2}$, where $\alpha_0$
describes the static disorder, and the second term is from the
dynamic structural fluctuations.

For the electrical transport measurement, at zero bias voltage
all the localized states are occupied, resulting
from the hybridization of the individual HOMO states of all the base pairs.
Switching on the bias voltage, a non-equilibrium state
is developed, which is described kinetically by the rate equation
\bea\label{KE-1}
\dot{f}_{n}&=&(1-f_{n})\sum_{m}k_{nm}f_{m}
            -\sum_{m}(1-f_{m})k_{mn}f_{n}  \nl
           && +k_{n}^{\rm in}(1-f_{n})-k_{n}^{\rm out}f_{n} \text{.}
\eea
Here $f_{n}$ is the probability of hole occupation.
In this equation, two types of hopping rates are involved, i.e.,
hopping between localized states in the DNA molecule,
and hopping between the (localized) molecular states and the electrodes.
The former has been given by the standard VRH model.
In the following we develop an expression for the latter,
which is characterized by the rates $k_{n}^{\rm in}$ and $k_{n}^{\rm out}$.

In contrast to quantum transport,
the {\it classical} hopping considered here involves
{\it real transition} between the electrode and the localized molecular
states with different energies, and the individual excess energy is
gained from or lost into the surrounding environment.
In general, this inelastic hopping process is described
by the non-radiative transition, with rate
\begin{equation}
k_{n\mathbf{p}}=\frac{2\pi }{\hbar }\vert H_{n\mathbf{p}}\vert ^{2}
         ( 4\pi\lambda k_{B}T) ^{-\frac{1}{2}}
         \exp\left[ -\frac{ ( E_{\mathbf{p}n}-\lambda ) ^{2}}
         {4\lambda k_{B}T} \right] .  \label{KE-3}
\end{equation}
Here, $\lambda $ is the reorganization energy of the environment,
and $E_{\mathbf{p}n}= \epsilon_{\mathbf{p}}-\epsilon_n$, denoting
the energy difference between the electrode state with momentum $\mathbf{p}$
and the $n$th localized molecular state,
which are coupled with strength $H_{n\mathbf{p}}$.
Physically, $H_{n\mathbf{p}}\simeq H\exp (-\beta R_n/2)$,
where $\beta^{-1}$ is the tunnelling length which is assumed here
independent of the electrode states,
and $R_n$ is the distance between the electrode and the $n$th localized state.
Also, in later numerical evaluation of the hopping rates
between the molecular states and the electrodes,
we would approximate the energy difference
$E_{\mathbf{p}n}\simeq\epsilon_{\mathbf{p}}-\epsilon_0$,
where $\epsilon_0$ denotes the energy of the localized molecular
state nearby the electrode.
This approximation makes sense in viewing the rapid decay of
$H_{\mathbf{p}n}$ with $R_n$.
The total hopping rate from the electrode to the localized molecular state
can be evaluated by integrating the electrode states.
For instance, the hopping rate from the left electrode to the
$n$th molecule state reads,
$k_{nL}=g_L\int d\epsilon_p k_{n\mathbf{p}} f_L(\epsilon_p-\mu_L)$,
where $g_L$ is the density of states of the (left) electrode.
In the following numerical calculation, the combined parameters
$\Gamma_{L(R)}=2\pi |H|^2 g_{L(R)}$
will be used to characterize the coupling strength between the molecule
and electrode, and will be commonly adopted as 0.2 meV.
The Fermi level of the left electrode in the presence of applied voltage $V$
is assumed to be $\mu_L=E_{g0}+eV/2$,
where the gap energy $E_{g0}=E_F-\epsilon_0$, with $E_F$ the equilibrium Fermi
energy of the electrode.
Other hopping rates between the electrode and the localized molecular
states, i.e., $k_{Ln}$, $k_{Rn}$ and $k_{nR}$, can be similarly evaluated.
Thus the rates in \Eq{KE-1} are obtained as
$k_{n}^{\rm in}= k_{nL}+k_{nR}$,
and $k_{n}^{\rm out}= k_{Ln}+k_{Rn}$.

After identifying all the rates in \Eq{KE-1},
we can obtain the time-dependent evolution of the occupation
probabilities on the individual localized states in response to
an applied voltage.
In this work we are in particular interested in the stationary hopping current
through the DNA molecule, which can be evaluated as
\bea
I=e\sum_n[(1-f_n)k_{nL}-k_{Ln}f_n],
\eea
under the condition $\dot{f}_n=0$.

\begin{figure}[h]
\begin{center}
\centerline{\includegraphics [scale=0.8] {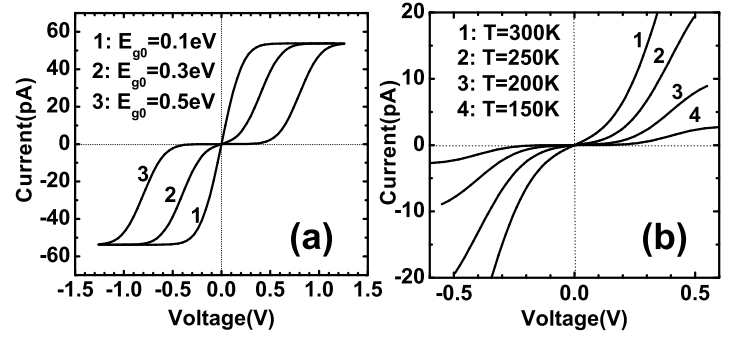}}
\end{center}
\caption{ I-V characteristics under hopping conduction
through DNA molecule with $N=30$ base pairs.
Plotted are results for,
(a) different energy gaps at temperature $T=300$ K,
and (b) different temperatures with a given $E_{g0}=0.3$ eV.
Other parameters adopted here are
the reorganization energy $\protect\lambda =0.1$ eV,
and the disorder energy $\Delta =0.15$ eV. }
\end{figure}

Figure 1 shows the I-V characteristics associated with the hopping conduction,
where the DNA molecule with 30 base pairs is exemplified.
Here the calculated current is eventually
saturated at certain bias voltage, as shown in Fig.\ 1(a),
owing to adoption of the simplified {\it one-band} model.
In this work we will use the saturated maximum current ($I_{\rm max}$)
to characterize the conduction ability (equivalent to the
average conductivity over different voltages).
The voltage gap in Fig.\ 1(a) is roughly determined by the
relative position of the HOMO level of the DNA base pair near the
electrode from the Fermi surface of the electrode at equilibrium,
i.e., $eV_{g}\approx 2|E_{g0}-\lambda |$. Note that this gap
differs from its counterpart in the {\it quantum} transport
regime, where the individual base-pair states interact with each
other and an energy band is formed, and the voltage gap is
determined by the distance of the upper edge of the energy band
from the electrode Fermi surface \cite{Hjo01,Li01,Dek02}.
In different experiments, this gap may differ considerably,
leading to either the metallic ohmic or the semiconductor behaviors.
We thus adopted several values of $E_{g0}$ in
Fig.\ 1(a) to illustrate the possibly observed I-V
characteristics.
Moreover, the hopping conduction displays a characteristic
temperature dependence of thermal activation, as shown in Fig.\ 1(b),
which is in good agreement with the experiment by Yoo {\it et al} \cite{Yoo01}.

\begin{figure}[h]
\begin{center}
\centerline{\includegraphics [scale=0.65] {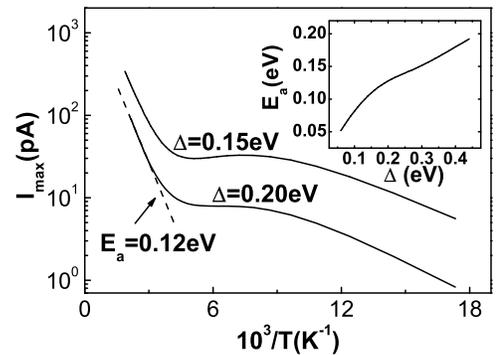}}
\end{center}
\caption{Maximum (saturated) current versus inverse temperature. The
adopted parameters are $\protect\lambda =0.1$ eV, $E_{g0}=0.3$ eV
and $N=30$. The inset shows the thermal activation energy $E_{a}$
versus the disorder strength $\Delta $.}
\end{figure}

The thermal activation characteristics are further
manifested clearly by the exact exponential dependence of
the inverse temperature at high temperature shown in Fig.\ 2,
where the slope gives the thermal activation energy $E_a$.
In general, the thermal activation energy depends on
the disorder, as quantitatively shown in the inset of Fig.\ 2.
As an illustration, for disorder $\Delta =0.2eV$, we obtain $E_{a}=0.12eV$,
which agrees well with both the experiment \cite{Yoo01}
and the {\it ab initio} calculation \cite{Ale03}.
Lowering the temperature, there appears a notable regime in which
the conduction is of weak dependence of the temperatures.
The transition takes place at temperature of
$200\sim 250$K, which is again in agreement with the
experiment \cite{Yoo01}.
The results numerically obtained here
can be qualitatively understood by the VRH argument \cite{Yu01}.

\begin{figure}[h]
\begin{center}
\centerline{\includegraphics [scale=0.65] {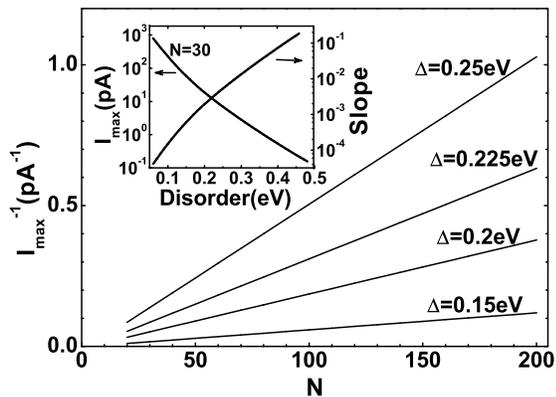}}
\end{center}
\caption{Ohmic behavior of length dependence under hopping conduction.
The plotted lines correspond to different disorder strengths.
Other parameters are $\lambda =0.1$ eV, $E_{g0}=0.3$ eV, and $T=300$ K.
The inset displays the effective resistivity (obtained from the
slope of the lines in the main part of the figure) versus the disorder strength,
and the conducting-to-insulating transition (i.e. with current from nA to pA)
by increasing the disorder for a DNA with $N=30$ base pairs. }
\end{figure}

Another characteristic feature associated with the hopping conduction
is the ohmic behavior of length dependence, as shown in Fig.\ 3.
This feature differs from either the
coherent tunnelling through a disordered system at zero
temperature, or the quantum transport through system without
disorder: the former has the characteristic length dependence
$\sim (e^{-2\alpha L}-1)$ \cite{And80},
while the latter leads to a maximum
current contributed from the entire HOMO band which is almost
independent of the molecule length.
The disorder strength would significantly affect the conduction property
as manifested in the inset of Fig.\ 3,
by the effective resistivity (i.e. the slope)
and the conducting-to-insulating transition (i.e. with current from nA to pA)
by increasing the disorder.
The insulating transition also happens by increasing the molecule length.
As a rough estimate, consider the hopping conduction through DNA molecule
at room temperature and with energy disorder
$\Delta=0.15$ eV. From the result in Fig.\ 3,
we obtain an estimate for the maximum current which would decrease from
60 pA to 0.5 pA as the base-pair numbers increase from 30 to 3000,
i.e., to the length of micron. This is nothing but the
insulating transition of DNA molecules on micron scales,
which has been commonly concluded in a number of recent experiments
\cite{Pab00,Dek01,Ong02}.

In summary, we have presented a kinetic study for the
transport properties of disordered DNA molecules,
based on Mott's variable-range hopping theory.
A number of unique features associated with the thermal activated
hopping mechanism were discussed with respect to either the already
known experimental results, or the possible future experiments.


\vspace{5ex}
{\it Acknowledgments.}
Support from the National Natural Science Foundation of China,
the Major State Basic Research Project No.\ G001CB3095 of China,
and the Research Grants Council of the Hong Kong Government
are gratefully acknowledged.



\clearpage

\end{document}